\documentclass[pre, twocolumn, floatfix]{revtex4}
\usepackage{graphicx}
\usepackage{subfigure}
\usepackage{amssymb}
\usepackage{amsmath}
\usepackage{epstopdf}
\usepackage{bm}
\newcommand{\bb}{\begin{equation}}
\newcommand{\en}{\end{equation}}
\newcommand{\sech}{\, \mbox{sech}}

\begin{document}

\title{Capillary wave dynamics on supported viscoelastic films:
Single and double layers}

\author{Mark L. Henle$^1$ and Alex J. Levine$^{1,2}$}
\affiliation{$^1$Department of Chemistry and Biochemistry,
University of California, Los Angeles, CA 90095\\
$^2$ California Nanosystems Institute, University of California,
Los Angeles, CA 90095}

\date{\today}

\begin{abstract}
We study the capillary wave dynamics of a single viscoelastic supported
film and of a double layer of immiscible viscoelastic supported
films. Using both simple scaling arguments  and a continuum
hydrodynamic theory, we investigate the effects of
viscoelasticity and interfacial slip on the relaxation dynamics
of these capillary waves. Our results account for the recent
observation of a wavelength-independent decay rate for
capillary waves in a supported polystyrene/brominated polystyrene
double layer [X. Hu {\em et al.}, Phys. Rev. E {\bf 74}, 010602 (R) (2006)].
\end{abstract}

\maketitle

\section{Introduction}

Due to the presence of thermally excited
capillary waves, the free surfaces of fluids~\cite{lamb:32}, complex fluids, 
and other generically soft materials are
fluctuating structures.  Examining the surface dynamics of such soft
materials provides a window into their rheology.
Consequently, capillary waves and interfacial dynamics
have been probed by light scattering
techniques~\cite{hard:87,hughes:93} in a variety of systems, including
membranes and monolayers~\cite{kramer:71}, liquid
metals~\cite{earnshaw:82}, and polymer
solutions~\cite{huang:96,munoz:05}, brushes~\cite{kim:05}, and
gels~\cite{langevin:98}.  These light scattering techniques probe
surface dynamics at length scales in the micron range; however,
newer experiments using x-ray photon correlation spectroscopy
(XPCS)~\cite{thurn:96,gutt:03} extend these measurements into the
sub-micron range, allowing investigators to probe even smaller scale
surface height fluctuations.

The theory of the surface dynamics of polymeric materials has focused
on the continuum mechanics of viscoelastic liquids, including two-fluid approaches to
semi-infinite polymer solutions~\cite{harden:91} 
and single component fluids supported on a
rigid substrate~\cite{jackle:98,rauscher:05, kumaran:92}.
These continuum approaches have been remarkably successful in
accounting for the surface dynamics observed experimentally in many systems. In
polymeric systems such continuum based
methods must begin to fail in the limit of increasing molecular
weight and decreasing film thickness, where the short dimension of
the liquid becomes comparable to size of the constituent molecules.
Indeed, there is experimental evidence~\cite{keddie:94} for and
subsequent theoretical conjecture~\cite{degennes:00,mccoy:02} about
the shift (relative to its bulk value) in the glass transition
temperature of high molecular weight polymer thin films.

In recent experiments Lal, Sinha, and coworkers~\cite{kim:03, lal:05}
used XPCS to study capillary waves on layers of polymeric films.
They examined both single layers of polystyrene (PS) supported on a
silicon substrate~\cite{kim:03} and double layers of a PS layer overlying
a brominated polystyrene (PBrS) layer on a silicon substrate~\cite{lal:05}.   
In the double layer experiments, the XPCS technique
allowed for the independent measurement of the dynamics of both the
upper surface of the PS, which we refer to as the free surface, and the PS/PBrS interface,
which we refer to as the buried interface.

The most striking result of the Lal \textit{et al.} double layer experiments is
the appearance of a slow ($\sim 100$ s) decay rate in the height-height
correlations of both the upper and buried interfaces
that is approximately \textit{independent of the in-plane wave number $q$}.
Such a phenomenon cannot be explained by previous
theoretical work~\cite{harden:91,jackle:98,rauscher:05, kumaran:92}.
In this article we extend the previous theoretical work to two-layer viscoelastic
systems and explore -- via continuum visco-hydrodynamic calculations
and scaling arguments -- the possible origins of the $q$-independent decay
rate reported by Lal and collaborators.  Acknowledging that the
thickness of the fluid layers is on the order of a few radii of
gyration of the constituent molecules, one may question whether
these discrepancies point towards the failure of a continuum-based
analysis. Despite this concern, we show that a viscoelastic
continuum model can give rise to the observed
$q$-independent decay rate. In order to quantitatively account for the
experimental data, however, we must use surprisingly long stress relaxation times in
these continuum models, which in turn leads us to postulate that the
polymer dynamics in the thin films is hindered by confinement effects. Based on these attempts to
fit the scattering data using our model, we suggest that the experiments provide an interesting
measure of confinement effects on the molecular dynamics in the melt.

The remainder of the paper is organized as follows:  in
section~\ref{sec:scaling} we examine simple scaling arguments that
suggest that viscoelastic supported films can exhibit the phenomena
reported by Lal and collaborators.  In section~\ref{sec:single} we calculate the
dynamics of the single supported fluid layer system using a continuum
hydrodynamic description of the fluid. We consider
both a purely viscous Newtonian fluid and the simplest model for a viscoelastic fluid,
the Maxwell fluid, which is characterized by a single relaxation time.
We also examine the effects of slip at the fluid/substrate
interface on these dynamics; in particular, we show that slip alone cannot
account for the $q$-independent decay rate. We then turn to the study of
the two layer system in  section~\ref{sec:double}. Here, we consider both
the system of two Newtonian fluids as well as the case of one Newtonian
fluid and one Maxwell fluid. We show that our model of a Maxwell fluid
buried beneath a Newtonian fluid can account for a number of
the experimental features observed in the double layer systems
studied by Lal \textit{et al.}. We again explore the effects of slip on the
dynamics of the system, in this case at the liquid/liquid interface.
Finally, in section~\ref{sec:conclusion}, we discuss our results more broadly in
the context of the dynamics of multilayered systems, focusing on the
long stress relaxation times necessary to account for the results of the
Lal group, as well as the implications such times have for the dynamics
of polymers near an interface. We conclude with  suggestions for future
experiments to further test our analysis.

\section{Scaling Analysis}
\label{sec:scaling}

\begin{figure}
\includegraphics[width=8.5cm]{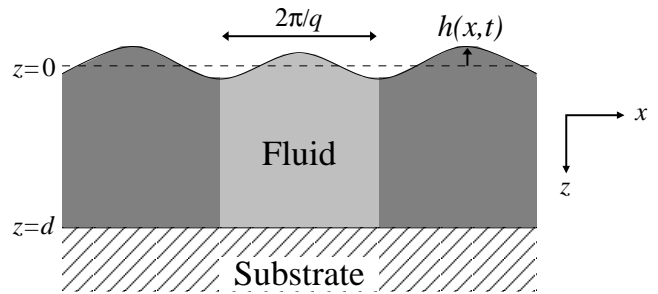}
\caption{\label{fig:SingleLayer}  Schematic illustration of a single
supported fluid layer of thickness $d$.}
\end{figure}

The experiments of Lal \textit{et al.}~\cite{kim:03, lal:05} present an
interesting theoretical challenge for which at least a few
suggestions have been offered~\cite{harden:91,jackle:98,rauscher:05}.
  We first use a few numerical
estimates to narrow the focus of our problem. Consider a supported
fluid film of thickness $d$ on a solid substrate,  as shown in
Fig.~\ref{fig:SingleLayer}.
The fluid has mass density $\rho$ and a surface
tension $\gamma$ at the free surface.  The deformation of the free
surface in capillary waves is subject to restoring forces due to
gravity and surface tension. We assume that the bending energies of the
interface are negligible (we will return briefly to this
point in section~\ref{sec:conclusion}).
The importance of these two forces depends strongly on the wavelength
of the disturbance. For wavelengths less than the capillary
length $\sqrt{\gamma/ \rho g}$, surface tension induced forces dominate
over gravitational forces. The XPCS experiments
can detect capillary waves of wavelengths less than the typical
transverse coherence length of the beam, which is $\sim 10 \mu$m
~\cite{kim:03,gutt:03, lal:05}. The capillary wavelength, however, is on
the order of $1$ cm; therefore, we may neglect the effect of gravity.

The importance of inertial effects in the fluid dynamics may be
estimated by considering the Reynolds number~\cite{landau:04} of the
flows associated with the capillary waves. The Reynolds number for
such flows is given by $\mbox{Re} \sim D_1 h \omega/\nu$, where $D_1$
is the length scale over which the fluid velocity vanishes (we
expect it to the be the lesser of the inverse wave number $1/q$ and
the film thickness $d$), $\nu = \eta/\rho$ is the kinematic viscosity
and $h$, $\omega$ are the typical height and decay rate
of the surface disturbance, respectively. Using the equipartition theorem to
determine the average magnitude of thermally generated capillary
waves, $\langle |h_q |^2 \rangle \sim k_{\rm B} T/\gamma q^2$, we
find that $\mbox{Re} \sim {\mathcal O}(10^{-7})$. This estimate demonstrates that inertial
stresses are greatly dominated by viscous ones in the material. Thus, we
ignore inertial stresses in the remainder of this article; that is, we assume that
all of the fluids are completely overdamped.

Given that we are considering low Reynolds number dynamics at scales
well below the capillary length, we now turn to simple scaling arguments
to determine the possible wave number dependence of the relaxation
rates of overdamped capillary waves. In particular, we ask: what
properties of the fluid lead to a
$q$-independent dispersion relation, \emph{i.e.} $\omega(q) \sim q^0$?
The fluid deformations generated by capillary waves store energy in
the fluid interface and, for viscoelastic fluids, in the bulk as well.
When these deformations relax, this energy is dissipated through
viscous stresses in the fluid. The scaling behavior of the dispersion relation
can then be found by equating the power generated by relaxing the elastic
deformations in the fluid with the power dissipated viscously by the fluid.

For a Newtonian fluid, energy can be stored only in the interface at
the free surface. Consider the lightly shaded portion of the fluid
in Fig.~\ref{fig:SingleLayer}, whose cross-sectional area in the
$x-y$ plane is $A \sim l/q$, $l$ being a unit length in  the
${\bf \hat{y}} $ direction.  The total power transferred to this volume of fluid
from the interface is the product of the  normal stress generated by
the surface tension $\gamma q^2 h$, the surface area $A$, and speed of the surface
 as the deformation relaxes, $\omega h$:
\begin{equation}
\label{eq:Psurf}
P_{surf} \sim \left(\gamma q^2 h\right)A \omega h.
\end{equation}

The viscous force density for the incompressible fluid is $\eta
\nabla^2 v$, where $v$ is the fluid velocity.  Then the power
dissipated in the bulk of the fluid is given by
\begin{equation}
\label{eq:Pdiss}
P_{diss} = \int d^3\!x
\, v  \eta \nabla^2 v \approx A \eta \int_0^d d\! z \, \bar{v} (z)
\nabla^2 \bar{v} (z),
\end{equation}
where $\bar{v} (z)$ is velocity averaged over one wavelength
in the horizontal (${\bf \hat{x}}$) direction; since the velocity is periodic in $x$, its
variation in $x$ does not affect the scaling behavior of the
dispersion relation.

We can relate the fluid velocity $\bar{v}$ to the interfacial height via
volume conservation. During the relaxation of a surface undulation, fluid volume
is transferred from elevated regions to the depressed ones, so  
$A \partial_t h \approx \int d\!y d\!z \bar{v} (z)$, where the integral is over a unit
of area whose normal is parallel to ${\bf \hat{x}}$.  Using
this relation and equating the power input to the power dissipated, we find
that the wave number dependence of the decay rate $\omega$ can be
determined from
\begin{equation}
\label{eq:omegaScale}
\omega \sim \frac{\gamma q^4\left[\int_0^d
d\!z \bar{v}(z)\right]^2}{\eta \int_0^d d\! z
\bar{v} (z) \nabla^2 \bar{v} (z)}.
\end{equation}

We now distinguish two different scaling regimes. In the thin layer
limit ($q d \ll 1$) the velocity decay into the fluid is set by
the thickness of the layer, so $\int_0^d d\!z \sim d$ and
$\nabla^2 \sim 1/d^2$. For thick layers ($q d \ll 1$), however, the
velocity decays exponentially into the fluid, so
$\int_0^d d\!z \sim 1/q$ and the $\nabla^2 \sim q^2$. Thus, we find
two possible dispersion relations depending on the the value of $q
d$,
\begin{equation}
\label{eq:omegaNewtScale} \omega \sim
\begin{cases}
\frac{\gamma q^4 d^3}{\eta} & q d \ll 1\\
\frac{\gamma q}{\eta} & q d \gg 1
\end{cases}.
\end{equation}
This argument suggests that a Newtonian fluid will not exhibit the
desired $q$-independent dispersion relation.

If we consider instead a viscoelastic fluid, we note that elastic energy is also stored 
in the bulk due to the state of deformation
in the material, as long as the decay rate of the capillary wave is
fast compared to the stress relaxation time in the viscoelastic
material. This storage of elastic energy in the bulk changes our
previous scaling arguments. For example, in an elastic solid
characterized by a bulk modulus $\mu$, the power released in the
bulk during the relaxation of the deformation takes the form
\begin{equation}
P_{bulk} \sim A \mu \int_0^d \bar{v}(z) \nabla^2 \bar{u}(z),
\end{equation}
where $\bar{u}(z)$ is the average displacement field in the medium:
$\bar{u}(z) \sim \bar{v}(z)/\omega$.  This expression is similar in
form to the viscous power dissipation in Eq.~(\ref{eq:Pdiss}).
Indeed, if we equate the power dissipated with the power input from
the bulk, we find $\omega \sim \mu/\eta$ for all values of $q$.  Thus, this
simple scaling argument suggests that the desired $q$-independent behavior
is a manifestation of the viscoelastic response of the fluid.  In order to
verify the results of these heuristic arguments, and to determine over which range
of wave numbers one might observe a  $q$-independent
dispersion relation, we now turn to the complete solutions of the Stokes
equation for supported, overdamped Newtonian and viscoelastic fluid layers.

\section{Single Layer}
\label{sec:single}

Consider the response of the single, supported, incompressible fluid layer shown
in Fig.~\ref{fig:SingleLayer} to an external stress field
$\sigma^{ext}$, which is normal to the free surface and characterized by frequency $\omega$
and wave vector ${\bf q} \equiv q {\bf \hat{x} }$,
\bb
\label{eq:SigmaExtSL}
\sigma^{ext} = \sigma_0 e^{i \left(q x-\omega t\right)}.
\en
The response of the fluid to
the applied stress is described by the vertical deviation or height
function of the free surface $h (x, t)$, the velocity
${\bf v}(x, z, t)$, and the pressure $P (x, z, t)$.  Given the
form of $\sigma^{ext}$ in Eq.~(\ref{eq:SigmaExtSL}), all of these
dynamical quantities can be written in the form
\begin{equation}
\label{eq:Transform} g (x,z,t) = g (q,
z,\omega) e^{i\left(q x- \omega t\right)}.
\end{equation}
The general solution to the Stokes equation in the case that the
dynamical quantities take the form of Eq.~(\ref{eq:Transform}) is
determined in Appendix~\ref{app:NavierStokes}. In particular, all of the
dynamical quantities listed above can be related to the normal component
of the fluid velocity, which is given by Eq.~(\ref{eq:VelocityNI}).
In order to determine the four integration constants in
Eq.~(\ref{eq:VelocityNI}), we need to specify the boundary
conditions at the top and bottom boundaries of the fluid.
At the fluid/substrate interface,
the normal component of the fluid velocity must vanish,
\bb
\label{eq:BC1SL}
\left.v_z \right|_{z=d}= 0.
\en
We may account for both slip and stick boundary conditions on the
tangential velocity component by introducing a
slip length $\lambda$, so that at the fluid/substrate interface
\bb
\label{eq:BC2SL}
\left.v_x \right|_{z=d} = -\lambda \left.\partial_z v_x
\right|_{z=d}.
\en
Taking $\lambda = 0$ reduces the above boundary condition to the usual
no-slip condition.

At the free surface, the rate of change of the height of the
interface must equal the fluid velocity at that point,
\begin{equation}
\label{eq:BC3SL}
\left.v_z \right|_{z=0} = -
i \omega h(q, \omega).
\end{equation}
Furthermore, this interface cannot support shear stresses,
\begin{equation}
\label{eq:BC4SL}
\left. \sigma_{xz}^{f} \right|_{z=0} = 0,
\end{equation}
where the fluid stress tensor $\sigma^f$ takes the usual form,
\begin{equation}
\label{eq:FluidStress}
\sigma_{ij}^{f} = \eta (\omega) \left(
\partial_i v_j + \partial_j v_i \right) - \delta_{ij} P,
\end{equation}
$\eta (\omega)$ being the frequency dependent
viscosity characterizing the viscoelastic
response of the material.  Finally, the free surface can
support a stress discontinuity between the externally applied stress
and the hydrodynamic stresses in the bulk material,
due to the presence of a finite surface tension:
\begin{equation}
\label{eq:StressBalanceSL1}
\sigma^{ext} = \gamma q^2 h (q, \omega)
- \left. \sigma_{zz}^{f} \right|_{z=0}.
\end{equation}

Using the boundary conditions Eqs.~(\ref{eq:BC1SL})-(\ref{eq:BC4SL})
to solve for the fluid velocity and pressure fields, we obtain the normal fluid
stress at the free surface from Eq.~(\ref{eq:FluidStress}),
\begin{equation}
\left.\sigma_{zz}^{f} \right|_{z=0} = i \omega \eta (\omega)
B(q) h(q, \omega),
\end{equation}
where
\begin{equation}
\label{eq:B}
B(q) = 4q \frac{\cosh^2(q d) +
q^2 d^2 + q \lambda\left(2 q d+ \sinh (2 q d)\right)}{\sinh(2 q d)-
2 q d+ 4 q \lambda \sinh^2 (q d)}.
\end{equation}
Then Eq.~(\ref{eq:StressBalanceSL1}) becomes
\begin{equation}
\label{eq:StressBalanceSL} \sigma_0 = h (q, \omega) \left[\gamma q^2
- i \omega \eta (\omega) B (q)\right].
\end{equation}

The normal mode frequencies are those which satisfy Eq.~(\ref{eq:StressBalanceSL})
in the absence of an external stress and with a non-zero fluid height:
\begin{equation}
\label{eq:NormalModesSL}
\gamma q^2 -  i \omega_n \eta( \omega_n) B
(q)= 0.
\end{equation}
For this overdamped system the frequencies $\omega_n$, given by the
solutions to Eq.~(\ref{eq:NormalModesSL}), all lie on the negative
imaginary axis. We refer to the norm of these complex numbers
as the decay rates of the system, $\tilde{\omega}_n = i
\omega_n$.

The experimentally measurable quantity for this system is the
intensity autocorrelation function, which is directly proportional
to the height-height correlation function $S (q, t) \propto
\left<h(q, t) h^* (q, 0) \right> $ (see Appendix~\ref{app:FDT} for
further details). Using Eq.~(\ref{eq:HeightCorr}), we find the
height-height correlation function to be
\begin{equation}
\label{eq:HeightCorrSL} S (q, t) = -\sum_{\omega_n}
\lim_{\omega \rightarrow \omega_n} \frac{\left(\omega-
\omega_n\right)   e^{- i \omega t}}{\omega\left[\gamma q^2 -  i
\omega \eta( \omega) B (q)\right]}.
\end{equation}
Thus, the surface height dynamics -- as parameterized by the
decay rates $ \tilde{\omega}_n$ -- can be extracted from the experimental
measurement of $S (q, t)$.  We now turn to the calculation of these
decay rates for single supported Newtonian and Maxwell fluids. In
light of the Lal \emph{et al.} data we pay particular attention to
the dependence of these decay rates on wave number $q$.

\subsection{Newtonian Fluid}

For Newtonian fluid the viscosity is real, positive, and independent
of frequency, $\eta (\omega) = \eta>0$. This is generally a good
approximation for small molecule
liquids and for viscoelastic materials on time scales much longer
than their typical stress relaxation times. As shown in
Appendix~\ref{app:NumFreq}, we expect only one decay rate, and indeed
there is only one root of Eq.~(\ref{eq:NormalModesSL}) in this case,
\begin{equation}
\label{eq:OmegaNewtSL} \tilde{\omega} (q) =\frac{\gamma q^2}{\eta
B(q)}.
\end{equation}
We note that, for the case where there is no slip between the fluid and
substrate, the same result can be obtained from the previous theoretical
work of J\"{a}ckle~\cite{jackle:98, kim:03}.

\begin{figure}
\includegraphics[width=8.5cm]{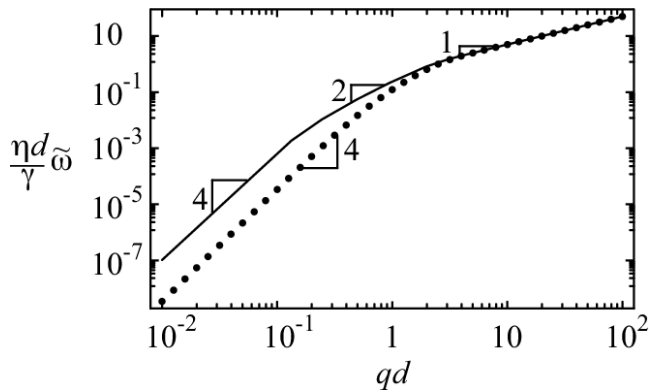}
\caption{\label{fig:OmegaNewtSL}Dimensionless decay rate $\eta
d \tilde{\omega} / \gamma$ for a Newtonian fluid as a function of $q
d$. The solid (dotted) lines are for a system with a small (large)
liquid/substrate slip length,  $\lambda = 0.01 d$ ($\lambda=10 d$).}
\end{figure}

The decay rates in the presence of both small and large slip lengths
are shown in Fig.~\ref{fig:OmegaNewtSL}. When the  slip length  is
small, $\lambda \ll d$, the decay rate is independent of it to
leading order in $q$,
\begin{equation}
\tilde{\omega} (q) =
\begin{cases}
\frac{\gamma q^4 d^3}{3 \eta}  & qd \ll 1\\
\frac{\gamma |q|} {2 \eta}& qd\gg 1\\
\end{cases}.
\end{equation}
Other than numerical prefactors, this result is identical to the one
obtained using the simple scaling arguments given above -- see
Eq.~(\ref{eq:omegaNewtScale}).  It is also consistent with the experimental 
results obtained by Sinha \textit{et. al} for a single PS layer,
for which a $q^4$ scaling behavior was observed over wave numbers
$q d <1$~\cite{kim:03}

When a significant amount of slip occurs between the fluid and the
solid interface (i.e. $\lambda \gg d$), an intermediate scaling
regime in the decay rate appears,
\begin{equation}
\tilde{\omega} (q) =
\begin{cases}
\frac{\gamma q^4 d^2 \lambda}{\eta} & qd \ll1, \, q^2d\lambda \ll 1\\
\frac{\gamma q^2 d}{4 \eta} & qd \ll 1, q^2 d \lambda \gg 1\\
\frac{\gamma |q|}{2 \eta} & qd \gg 1
\end{cases}.
\end{equation}
Thus, we can see that a Newtonian
fluid does not exhibit a $q$-independent decay rate, even if there is a
significant amount of slip between the fluid and substrate.

Because there is only one decay rate, the height-height
correlation function Eq.~(\ref{eq:HeightCorrSL}) exhibits a simple
exponential decay,
\begin{equation}
\label{eq:HeightCorrNewtSL}
S (q, t) = \frac{1}{\gamma q^2} e^{-\tilde{\omega} t}.
\end{equation}

\subsection{Maxwell Fluid}

A Maxwell fluid, which is the simplest model for a viscoelastic
material, has a complex, frequency dependent viscosity of the form
\begin{equation}
\label{eq:EtaMaxwell}
\eta (\omega) = \eta + \frac{E \tau}{1- i
\omega \tau},
\end{equation}
where $E$ is the transient modulus of the polymer network, $\tau$
is the stress relaxation time of the medium, and $\eta$ accounts
for the high frequency viscous response~\cite{landau:elasticity}. In this case
 Eq.~(\ref{eq:NormalModesSL}) can be written as a quadratic equation in $\omega$;
 that is, we obtain two decay rates, in agreement with the
arguments given in Appendix~\ref{app:NumFreq}. These rates are given by
\begin{equation}
\label{eq:OmegaMaxwellSL}
\tilde{\omega} _\pm (q) =
\frac{\mathcal{N}\pm \sqrt{\mathcal{N}^2-4 \beta \eta \tau}}{2
\eta \tau},
\end{equation}
where $\mathcal{N} \equiv \eta+ E \tau +\beta \tau$ and  $\beta
(q) =  \gamma q^2 /B(q)$. Since $\mathcal{N}^2-4 \beta \eta \tau=
\left(\eta+ E \tau-\beta \tau\right)^2+4 \beta E \tau^2>0$ both
the fast decay rate $\tilde{\omega} _+$ and slow decay rate
$\tilde{\omega} _-$ are positive real numbers.

\begin{table}

\renewcommand{\arraystretch}{1.25}

\begin{tabular}{|c|c|c|}
\hline
Parameter & symbol & Value  \\\hline\hline
PS layer thickness & $d_2$ & $100$ nm \\
PBrS layer thickness & $d$ [SL], $d_1$ [DL] & $200$ nm \\
PS viscosity & $\eta_2$ & $10^4$  kg/m$\cdot$s \\
PBrS viscosity & $\eta$ [SL], $\eta_1$ [DL] & $10^6$ kg/m$\cdot$s \\
PBrS surface tension [SL] & $\gamma$ &  $10^{-2}$ N/m \\
Interface tension PS/PBrS & $\gamma_1$ & $10^{-3}$ N/m \\
PS surface tension & $\gamma_2$ &  $10^{-2}$ N/m \\
PBrS Plateau modulus & $E$ & $10^3$ Pa \\
PBrS stress relaxation time & $\tau$ & $100$ s \\ \hline
\end{tabular}

\caption{\label{tab:values} Geometric and rheological parameters
corresponding to the single layer [SL] and double layer [DL]
 Lal experiments \cite{lal:05}. Both the plateau
modulus and the stress relaxation time are chosen so as to reproduce
the major features observed experimentally in the double layer system
(see section~\ref{sec:double}).}
\end{table}

\begin{figure}
\includegraphics[width=8.5cm]{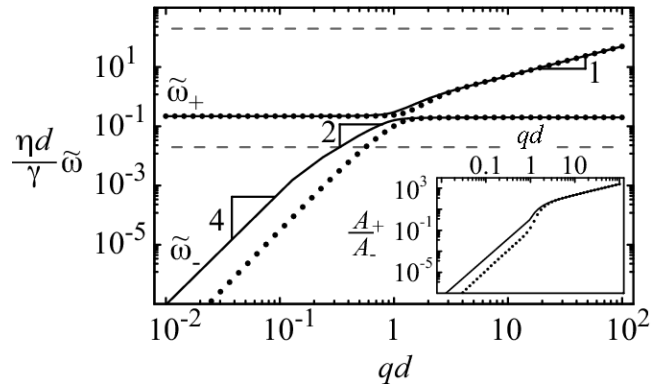}
\caption{\label{fig:MaxwellSL}Dimensionless decay rates
$\eta d \tilde{\omega}_\pm / \gamma$ for a Maxwell fluid as a
function of $q d$, using the parameter
values listed in Table~\ref{tab:values} for PBrS.  The solid (dotted) lines 
are for a system with a small (large)
liquid/substrate slip length,  $\lambda = 0.01 d$ ($\lambda=10 d$).  The dashed lines indicate
the approximate window of decay rates that can be measured in the
experiments, $10^{-3} s^{-1} < \tilde{\omega}<10 s^{-1}$. \textit{Inset}:  Ratio of
the amplitude of the term in $S (q,t)$ that decays with rate
 $\tilde{\omega}_+$ to that of the term that decays with
rate $\tilde{\omega}_-$.}
\end{figure}

In Fig.~\ref{fig:MaxwellSL} we plot the fast and slow decay rates
for both small and large amounts of slip at the substrate, using parameter values
characteristic of a layer of PBrS, as listed in Table~\ref{tab:values}.  When the
slip length is small compared to the film thickness, $\lambda \ll d$,
we find
\begin{align}
\label{eq:OmegaMaxwellSL1} \tilde{\omega}_+ (q) &=
\begin{cases}
\frac{\eta + E \tau}{\eta \tau} & q d \ll 1\\
\frac{\gamma |q|}{2 \eta} & q d \gg 1
\end{cases},\\
\notag \\
\label{eq:OmegaMaxwellSL2}
\tilde{\omega}_- (q) &=
\begin{cases}
\frac{\gamma q^4 d^3}{3\left(\eta + E \tau\right)} & q d \ll 1\\
\frac{1}{\tau} & q d \gg 1
\end{cases}.
\end{align}
When the slip length is large compared to film thickness, $\lambda
\gg d$, we find
\begin{align}
\label{eq:OmegaMaxwellSL3}
\tilde{\omega}_+ (q) &=
\begin{cases}
\frac{\eta + E \tau}{\eta \tau} & q d \ll 1\\
\frac{\gamma |q|}{2 \eta} & q d \gg 1
\end{cases},\\
\notag\\
\label{eq:OmegaMaxwellSL4}
\tilde{\omega}_-  (q) &=
\begin{cases}
\frac{\gamma q^4 d^2 \lambda}{\eta + E \tau} & q d \ll 1, \, q^2 d \lambda \ll 1\\
\frac{\gamma q^2 d}{4\left(\eta + E \tau\right)} & q d \ll 1, \, q^2 d \lambda \gg 1\\
\frac{1}{\tau} & q d \gg 1
\end{cases}.
\end{align}
As we can see from both the analytic expressions and the numerical results
shown in Fig.~\ref{fig:MaxwellSL},  the fast (slow) decay rate is independent of
$q$ at low (high) wave numbers. Such behavior is in agreement with
the experimental results of Lal \textit{et al.}, and it is also consistent with the
scaling arguments given in section~\ref{sec:scaling} for the elastic response
of a viscoelastic fluid.  On the
other hand, the scaling of the fast and slow decay rates in
their respective $q$-dependent regimes -- at high $qd$ in the former
case, and at low $qd$ in the latter -- is identical to that of a
Newtonian fluid in these regimes.   Thus, the viscous response of
the fluid determines the scaling of $\tilde{\omega}_\pm$ where each
decay rate exhibits a strong dependence on the wavelength, whereas
the elastic response of the fluid gives rise to the $q$-independent
behavior of $\tilde{\omega}_\pm$.

From this analysis we predict that the height-height correlation
function $S (q, t)$ will exhibit a double exponential decay
for a Maxwell fluid.  In particular, Eq.~(\ref{eq:HeightCorr}) becomes
\begin{align}
\label{S-amps}
S (q, t) =& \sum_\pm \left[\frac{\left(\tilde{\omega}_\pm
\tau -1\right) }{\eta \tau B(q)  \tilde{\omega}_\pm \left(
\tilde{\omega}_\pm - \tilde{\omega}_\mp \right)}\right]e^{-
\tilde{\omega}_\pm t}
\\
\notag
\equiv & \sum_\pm A_\pm e^{- \tilde{\omega}_\pm t}.
\end{align}
These amplitudes are plotted in the inset of Fig.~\ref{fig:MaxwellSL}.

In order to predict the form of experimental measurements of $S (q, t)$,
one must consider the values
of both the decay rates and the $q$-dependent amplitudes appearing
in Eq.~(\ref{S-amps}).  In principle, the decay of the height-height correlation
function  $S(q, t)$ should always have the double exponential form
given in  Eq.~(\ref{S-amps}).  It is clear from the inset of Fig.~\ref{fig:MaxwellSL},
however, that the amplitude of one of these terms can dominate the other at
certain wave numbers.  In particular, the amplitude of the slow mode
dominates for $q d \ll1$, whereas the amplitude of the fast mode dominates
for $q d \gg 1$. In these regimes, it will be hard to measure both decay rates.
The full double exponential decay should be observable only in the
crossover region between these two regimes, which occurs at $q d \sim 1$
for parameter values consistent with current experiments.

We can see from the inset of Fig.~\ref{fig:MaxwellSL} that for both
$q d \ll 1$ and $q d \gg 1$, the amplitude of the $q$-dependent rate is much larger
than that of the $q$-independent rate.  Taken in isolation, this fact suggests that
it would be difficult to observe the $q$-independent rate in these regions.  In
predicting the experimentally observed behavior of $S (q, t)$,
however, it is important to take the value of the
decay rates themselves into account.  In particular, there is
a finite range of decay rates that can be measured:
the decay times can be too fast or too slow to be observed within the time
scales of the experiment.  We have indicated a reasonable experimental window --
from $1/10$ sec to $1000$ sec for the decay times -- by the dashed lines in
Fig~\ref{fig:MaxwellSL}.  Then we see, for example, that the slow decay rate becomes
too slow to be measured experimentally at long wavelengths.  As a result, the
faster $q$-independent decay rate could be observed in this regime, despite the fact that
its amplitude is much smaller than the slow decay rate. We do not comment on the sensitivity 
of the experiments to small amplitude capillary dynamics. If the larger amplitude mode is too slow and the 
smaller amplitude mode generates too small a surface height undulation, no interfacial dynamics may 
be detected.

This analysis suggests that for $q d < 1$ it should in principle be possible to
observe a $q$-independent decay of $S(q,t)$ over at least one decade of $q d$, depending on 
the sensitivity of the measurement. For higher
wave numbers, $q d > 1$, however, we expect this $q$-independent behavior to be obscured by the slower
$q$-dependent mode with a larger amplitude, since it will then be fast enough to be measured by the
current experiments.

\section{Double Layer}
\label{sec:double}

\begin{figure}
\includegraphics[width=8.5cm]{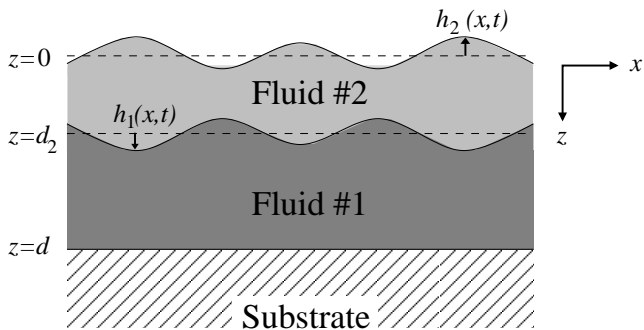}
\caption{ \label{fig:DoubleLayer}Schematic illustration of two fluid
layers, of thicknesses $d_1$ and $d_2$, on a substrate. The
coordinates are chosen so that the fluid/solid interface is the
plane $z=d$ and the unperturbed free surface is the plane
$z=0$.}
\end{figure}

We now turn to the double layer case, illustrated in
Fig.~\ref{fig:DoubleLayer}.  The buried and upper fluids have
viscosities $\eta_1 (\omega), \eta_2 (\omega)$ and thicknesses
$d_1, d_2$, respectively. The total thickness is $d = d_1+d_2$.
In general, we can drive this system by spatially oscillatory normal
stresses acting on both the buried fluid/fluid interface and the free
surface. As in the single layer case, we consider stresses of
the form of Eq.~(\ref{eq:SigmaExtSL}). Combining
the stresses on both interfaces into a single vector, we write
\begin{equation}
\label{eq:SigmaExtDL}
\bm{\sigma^{ext}} (x,t) =
\begin{bmatrix}
\sigma_1\\
\sigma_2\\
\end{bmatrix}
e^{i \left(q x-\omega t\right)}.
\end{equation}
For simplicity, we consider the possibility of slip at only the fluid/fluid
interface, as the effects of slip at the fluid/substrate interface have
already been explored in the single layer case.

At the fluid/substrate interface all components of the fluid velocity
must vanish:
\begin{equation}
\label{eq:BC1DL}
\left.v_{1,z} \right|_{z=d} =\left.v_{1,x} \right|_{z=d}= 0.
\end{equation}
At the fluid/fluid interface the $z$-components of the velocities of
the two fluids must match, but the tangential components of the
velocities do not if there is a finite slip length $\lambda$ at this
interface:
\begin{eqnarray}
\label{eq:BC2DL} \left. v_{1,z}\right|_{z=d_2} &=&
\left.v_{2,z}\right|_{z=d_2}, \\
\left. \left(v_{2,x} - v_{1,x}\right) \right|_{z=d_2} &=& - \lambda \left.
\left( \partial_z v_{2,x} - \partial_z v_{1,x} \right) \right|_{z=d_2}.
\end{eqnarray}
Note that $\lambda>0$ when the buried polymer layer is more viscous
than the upper layer ($\mbox{Re}\left[\eta_1 (\omega)\right] > \mbox{Re}
\left[\eta_2 (\omega) \right]$ for all real $\omega$), and vice-versa.

From the definition of the interface heights,
\begin{eqnarray}
\label{eq:BC3DL} \left. v_{1, z}\right|_{z=d_2} &=& -i \omega h_1
(q, \omega),\\
\left. v_{2, z}\right|_{z=0} &=& -i \omega h_2 (q, \omega).
\end{eqnarray}
In addition, the free surface cannot support shear
stresses, and the shear stress must be continuous across
the fluid/fluid interface:
\begin{eqnarray}
\label{eq:BC4DL} \left.\sigma_{2, xz}^f \right|_{z=0} &=& 0,\\
\left.\sigma_{1,xz}^f\right|_{z=d_2} &=& \left.
\sigma_{2,xz}^f\right|_{z=d_2}.
\label{eq:BC5DL}
\end{eqnarray}

Finally, the normal stress discontinuities at both the fluid/fluid
interface and the free surface are determined by their respective
surface tensions ($\gamma_1$ and $\gamma_2$, respectively),
\begin{eqnarray}
\label{eq:StressBalanceDL1} \sigma_1 &=& \gamma_1 q^2 h_1 (q,
\omega) -
\left[\sigma_{2,zz}^f-\sigma_{1,zz}^f\right|_{z=d_2},\\
\sigma_2 &=&  \gamma_2 q^2 h_2 (q,
\omega)-\left.\sigma_{2,zz}^f\right|_{z=0}.
\label{eq:StressBalanceDL2}
\end{eqnarray}
Using Eqs.~(\ref{eq:BC1DL})--~(\ref{eq:BC5DL}) to eliminate the
fluid velocities in favor of the interfacial height functions,
the normal stress equations Eqs.~(\ref{eq:StressBalanceDL1}) and
~(\ref{eq:StressBalanceDL2}) can be written in matrix form:
\begin{equation}
\bm{\sigma} = \bm{\Sigma} \cdot \bm{h}, \en
where
$\bm{\sigma} \equiv \begin{bmatrix} \sigma_1 \\ \sigma_2 \\
\end{bmatrix}$, $\bm{h} \equiv \begin{bmatrix} h_1 (q, \omega) \\
h_2 (q, \omega) \\ \end{bmatrix}$, and \bb \label{eq:StressMatrix}
\bm{\Sigma} \equiv
\begin{bmatrix}
\gamma_1 q^2-\frac{i \omega s_{11} (q, \omega)}{\Delta (q, \omega)} &
-\frac{i \omega s_{12} (q, \omega)}{\Delta (q, \omega)} \\
-\frac{i \omega s_{12} (q, \omega)}{\Delta (q, \omega)} & \gamma_2
q^2-\frac{i \omega s_{22} (q, \omega)}{\Delta (q, \omega)} \\
\end{bmatrix},
\end{equation}
with
\begin{widetext}
\begin{align}
\label{eq:s11}
\notag
s_{11} (q, \omega) =& 4 q \eta_1^2\left[1+\left(1+2 q^2 d_1^2\right)
\sech \left(2 q d_1\right)\right]\left[\tanh \left(2 q d_2\right)-2 q d_2 \sech
\left(2 q d_2\right) \right]+4 q \eta_2^2\left[1-\left(1+2 q^2 d_1^2\right)\sech
\left(2 q d_1\right)\right]\\
\notag
&*\left[\tanh \left(2 q d_2\right)+2 q d_2 \sech \left(2 q d_2\right) \right] +
8 q \eta_1 \eta_2 \left[2 q d_1\left(2 q^2d_1 d_2-1\right) \sech \left(2 q d_1\right)
\sech \left(2 q d_2\right)+\tanh\left(2 q d_1\right)\right]\\
&+8 q^2 \lambda \left(\eta_1-\eta_2\right)\Big\{\eta_1\left[1+\left(1+2 q^2 d_1^2\right)
\sech \left(2 q d_1\right)\right]\left[1-\sech \left(2 q d_2\right)\right]\Big.\\
\notag
&\Big.+\eta_2\left[\tanh \left(2 q d_1\right)-2 q d_1\sech \left(2 q d_1\right) \right]
\left[\tanh \left(2 q d_2\right)+2 q d_2 \sech \left(2 q d_2\right) \right]\Big\},
\\
\notag \\
\label{eq:s12}
\notag
s_{12} (q, \omega) =&- \frac{8 q \eta_2}{2 \cosh \left(q d_2\right)-\sech\left(q d_2\right)}
\Big\{ \eta_2\left[1-\left(1+2 q^2 d_1^2\right) \sech \left(2 q d_1\right)\right]
\left[q d_2+ \tanh \left(q d_2\right)\right] \Big. \\
&+\eta_1 \Big[2 q^3 d_1^2 d_2 \sech \left(2 q d_1\right)+\left[\tanh
\left(2 q d_1\right)- 2 q d_1 \sech \left(2 q d_1\right)\right]\left[q d_2 \tanh
\left( q d_2\right) + 1\right]\Big]\\
\notag
&+\Big.2 q \lambda\left(\eta_1-\eta_2\right)\left[\tanh \left(2 q d_1\right)- 2 q d_1
\sech \left(2 q d_1\right)\right] \left[\tanh \left( q d_2\right) + q d_2 \right]\Big\},
\\
\notag \\
\label{eq:s22}
\notag
s_{22} (q, \omega) =& 4 q \eta_2^2\left[1-\left(1+2 q^2 d_1^2\right) \sech
\left(2 q d_1\right)\right] \left[\tanh \left(2 q d_2\right)+2 q d_2 \sec\left( 2 q d_2\right)\right]\\
&+4 q \eta_1 \eta_2\left[\tanh \left(2 q d_1\right)-
2 q d_1\sech\left(2 q d_1\right)\right]\left[1+\left(1+2 q^2 d_2^2\right) \sech \left(2 q d_2\right)\right]\\
\notag
&+8 q^2 \lambda \eta_2 \left(\eta_1-\eta_2\right)\left[\tanh
\left(2 q d_1\right)-2 q d_1\sech \left(2 q d_1\right) \right]
\left[\tanh \left(2 q d_2\right)+2 q d_2 \sech \left(2 q d_2\right)\right],
\\
\notag \\
\label{eq:Delta}
\Delta (q, \omega) =& 2 \eta_2\left[1-\left(1+2 q^2 d_1^2\right)\sech
\left(2 q d_1\right)\right] \left[1-\sech \left(2 q d_2 \right)\right] +
2 \eta_1 \left[\tanh \left(2 q d_1\right)-2 q d_1\sech \left(2 q d_1\right)\right] \\
\notag &*\left[\tanh \left(2 q d_2\right)-2 q d_2\sech \left(2 q d_2\right)\right] +
4 q \lambda \left(\eta_1-\eta_2\right) \left[\tanh \left(2 q d_1\right)-
2 q d_1\sech \left(2 q d_1\right) \right]\left[1-\sech \left(2 q d_2 \right)\right],
\end{align}
\end{widetext}
where, for clarity, we have suppressed the (possible) frequency
dependence of the viscosities $\eta_1$ and $\eta_2$.

The two normal modes of the double layer system are easily
identified if we diagonalize the normal stress matrix equation
Eq.~(\ref{eq:StressMatrix}), bringing the dynamical relations into
the form
\begin{equation}
\label{eq:StressMatrixDiag}
\begin{bmatrix}
\sigma_+ \\
\sigma_-\\
\end{bmatrix}
=
\begin{bmatrix}
\lambda_+ & 0 \\
0 & \lambda_-\\
\end{bmatrix}
\begin{bmatrix}
h_+ (q, \omega)\\
h_- (q, \omega)\\
\end{bmatrix},
\end{equation}
where
\begin{equation}
\label{eq:SigmaPMhPM}
\sigma_\pm \equiv \Lambda_1^\pm \sigma_1 + \Lambda_2^\pm \sigma_2, \qquad
h_\pm \equiv \Lambda_1^\pm h_1+ \Lambda_2^\pm h_2,
\end{equation}
and $\lambda_\pm$ and $\Lambda^\pm$ are, respectively, the
eigenvalues and the orthonormal eigenvectors of the matrix
$\bm{\Sigma}$. The eigenvalues may be written as
\begin{equation}
\label{eq:Eigenvalues} \lambda_\pm = \frac{1}{2}\left[\mbox{Tr }
\bm{\Sigma} \pm \sqrt{\left(\mbox{Tr } \bm{\Sigma}\right)^2- 4 \det
\bm{\Sigma}}\right],
\end{equation}
and the corresponding eigenvectors are
\begin{equation}
\label{eq:Eigenvectors}
\bm{\Lambda^\pm} =
\frac{1}{\sqrt{\Sigma_{12}^2 + \left(\lambda_\pm -
\Sigma_{11}\right)^2}}
\begin{bmatrix}
\Sigma_{12} \\
\lambda_\pm - \Sigma_{11}\\
\end{bmatrix}.
\end{equation}
We can see from Eq.~(\ref{eq:StressMatrixDiag}) that $h_+$ and $h_-$
are the amplitudes of the two independent normal
modes of this double layer system.  As in the single layer case, the
characteristic decay rates of these two modes can be found by
solving the system of equations in the absence of external forces,
$\bm{\sigma} = {\bf 0}$. From Eq.~(\ref{eq:StressMatrixDiag}) we
note that the characteristic decay rates $\tilde{\omega}^\pm_n$  of
the $h_\pm$ mode are the roots of the eigenvalue $\lambda_\pm$, where
$n$ indexes the roots.

Examining Eq.~(\ref{eq:Eigenvalues}), it is clear that these roots are also
roots of the determinant of $\bm{\Sigma}$, which may be written as
\begin{widetext}
\begin{equation}
\label{eq:DetSigma}
\det \bm{\Sigma} (\omega) = \frac{\Phi (q, \omega)}{\Delta (q, \omega)}
(i \omega)^2-\frac{q^2}{\Delta (q, \omega)} \left[\gamma_1 s_{22}
(q, \omega) + \gamma_2 s_{11} (q, \omega)\right] i \omega + \gamma_1 \gamma_2 q^4,
\end{equation}
where
\begin{align}
\label{eq:Phi}
\notag
\Phi (q, \omega) =& 16 q^2 \eta_1 \eta_2^2\left[4 q^2 d_1 d_2
\left(1-q^2 d_1 d_2\right)\sech\left(2 q d_1\right) \sech
\left(2 q d_2\right)+\tanh\left(2 q d_1\right) \tanh \left(2 q d_2\right)\right]\\
\notag
&+8 q^2 \eta_2^3 \left[1-\left(1+2 q^2 d_1^2\right) \sech
\left(2 q d_1\right)\right] \left[1-\left(1+2 q^2 d_2^2\right) \sech
\left(2 q d_2\right)\right] \\
\notag
&+8 q^2 \eta_1 \eta_2^2 \left[1+\left(1+2 q^2 d_1^2\right) \sech
\left(2 q d_1\right)\right] \left[1+\left(1+2 q^2 d_2^2\right) \sech
\left(2 q d_2\right)\right] \\
&+16 q^3 \lambda \eta_2 \left(\eta_1-\eta_2\right)
\Big\{ \eta_1\left[\tanh\left(2 q d_2\right)+2 q d_2
\sech \left(2 q d_2\right)\right]\left[1+\left(1+2 q^2 d_2^2\right)
\sech \left(2 q d_2\right)\right]\Big. \\
\notag
&+\Big. \eta_2\left[\tanh\left(2 q d_1\right)-2 q d_1 \sech \left(2 q d_1\right)\right]
\left[1-\left(1+2 q^2 d_2^2\right) \sech \left(2 q d_2\right)\right] \Big\}
\end{align}
\end{widetext}
Furthermore, we can see from Eq.~(\ref{eq:Eigenvalues}) that the trace of $\bm{\Sigma}$
is positive (negative) for the roots $\tilde{\omega}^-$ ($\tilde{\omega}^+$).

Given the normal modes of the system, we can again use the results
of Appendix~\ref{app:FDT} to calculate the height-height correlation
functions $S_k (q, t) \propto \left<h_k (q, t) h_k^* (q,0)\right>$
for $k=1,2$, where $k=1$ labels the buried interface and $k=2$ the free surface.
Eq.~(\ref{eq:HeightCorr}) gives the normal mode correlation functions $S_\pm (q, t)$,
which can then be used to compute $S_k (q, t)$ using Eq.~(\ref{eq:SigmaPMhPM}),
\begin{align}
\label{eq:HeightCorrDL}
\notag
S_k (q, t) =& -\sum_{\omega_n}\frac{\Sigma_{\bar{k} \bar{k}}
(\omega_n) }{i \omega_n} \left[\frac{\partial}{\partial (i \omega)}
\left(\det \bm{\Sigma}\right)\right]^{-1}_{\omega=\omega_n} e^{-i \omega_n t} \\
& \equiv \sum_{\omega_n} A_n^{(k)} e^{- \tilde{\omega}_n t}
\end{align}
where $\bar{k} = 2, 1$ for $k=1,2$.

Clearly for the two layer system our solutions depend on a
larger set of geometric and material parameters. We do not show all
possible parameter regimes, but consider in both the following
figures and asymptotic results a parameter regime
consistent with the experiments of Lal \emph{et al.}~\cite{lal:05}. In particular, we take the
layer depths to be of the same order, $d_1 \sim d_2$, and the
surface tension of the fluid/fluid interface to be less than that of
the free surface, $\gamma_1 < \gamma_2$. We also assume that the
buried polymer layer is much more viscous than the upper layer,
$\mbox{Re} \left[\eta_1 (\omega) \right]\gg \mbox{Re}\left[ \eta_2 (\omega)\right]$ for all real $\omega$.
However, many of our results below (e.g. the scaling behavior of the normal
mode decay rates) apply more generally to two-layer viscoelastic systems.

\subsection{Two Newtonian Fluids}
\label{two-fluid:none-max}

\begin{figure}
\includegraphics[width=8.5cm]{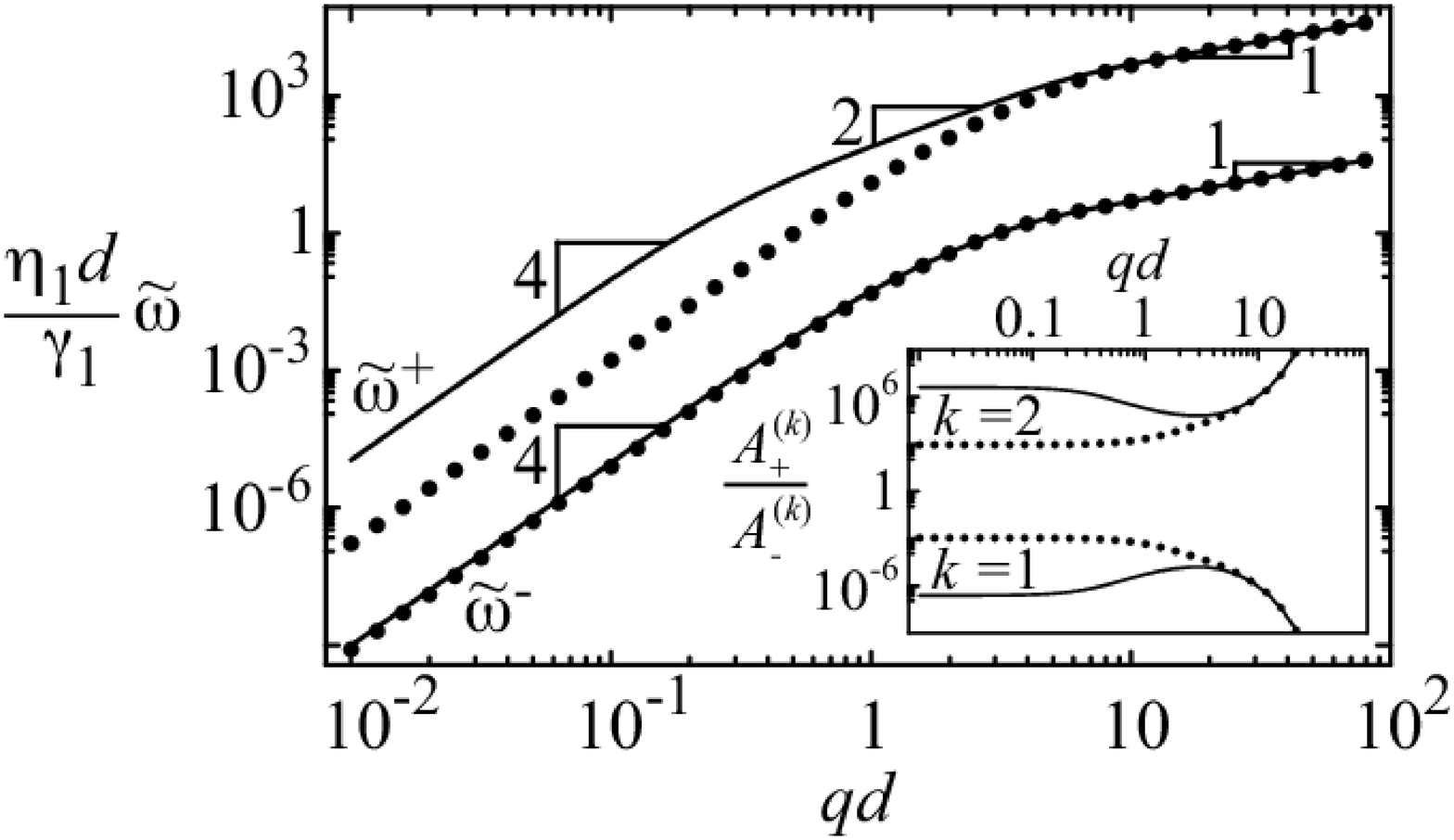}
\caption{\label{fig:NewtDL} Dimensionless decay rates $\eta_1
d \tilde{\omega}^\pm /\gamma_1$ and amplitude ratios $A_+^{(k)} /
A_+^{(k)}$ (\textit{inset}) as a function of $q d$ for a double layer system
with a buried Newtonian PBrS layer and an upper Newtonian PS layer,
using the parameter values given in Table~\ref{tab:values}.
The solid (dotted) lines are for a system with a small (large)
liquid/liquid slip length,  $\lambda = 0.01 d$ ($\lambda=10 d$).}
\end{figure}

In the case that both layers can be described as Newtonian fluids
with viscosities $\eta_1 (\omega)=\eta_1$ and $\eta_2 (\omega)=\eta_2$,
 it is clear from Eq.~(\ref{eq:DetSigma}) that $\det \bm{\Sigma}$ is a simple
quadratic function of $\omega$. Thus, there are two normal mode decay rates,
 as expected from the arguments given in
Appendix~\ref{app:NumFreq}. Specifically, the roots of
Eq.~(\ref{eq:DetSigma}) are given by
\begin{align}
\tilde{\omega}^\pm (q) =& \frac{q^2}{2 \Phi}
\Bigg[\gamma_1 s_{22}+\gamma_2 s_{11}\Bigg.\\
\notag
&\Bigg.\pm \sqrt{\left(\gamma_1 s_{22}+
\gamma_2 s_{11}\right)^2-4 \gamma_1 \gamma_2 \Phi \Delta}\Bigg]
\end{align}
where $\tilde{\omega}^\pm \equiv i \omega^\pm$.  The values of
the fast mode $\tilde{\omega}^+$ and the slow mode $\tilde{\omega}^-$ for
parameters consistent with a PBrS buried fluid and a PS upper fluid
(see Table~\ref{tab:values}) are plotted in Fig.~\ref{fig:NewtDL}. The $\tilde{\omega}^+$, 
$\tilde{\omega}^-$ modes correspond to nearly in phase and out of phase undulations of the
two surfaces, respectively.

Using Eqs.~(\ref{eq:s11})--(\ref{eq:Delta}), one can show that both
normal mode decay rates exhibit the same asymptotic scaling behavior
observed for the single viscous layer discussed above.
When the slip length is small, $\lambda \ll d$,
\begin{align}
\tilde{\omega}^+ (q) =&
\begin{cases}
\frac{\gamma_2 q^4 d_2^3}{3 \eta_2}  & qd \ll 1\\
\frac{\gamma_2 |q|} {2 \eta_2}& qd\gg 1\\
\end{cases} ,\\
\notag\\
\tilde{\omega}^- (q) =&
\begin{cases}
\frac{\gamma_1 q^4 d_1^3}{3 \eta_1}  & qd \ll 1\\
\frac{\gamma_1 |q|} {2 \eta_1}& qd\gg 1\\
\end{cases}.
\end{align}
A large slip length at the fluid fluid interface (i.e. $\lambda \gg
d$) produces an intermediate scaling regime in $\tilde{\omega}^+$,
\begin{equation}
\tilde{\omega}^+ (q) =
\begin{cases}
\frac{\gamma_2 q^4 d_2^2 \lambda}{\eta_2} & qd \ll1, \, q^2 d\lambda \ll 1\\
\frac{\gamma_2 q^2 d_2}{4 \eta_2} & qd \ll 1, q^2 d \lambda \gg 1\\
\frac{\gamma_2 |q|}{2 \eta_2} & qd \gg 1
\end{cases}.
\end{equation}

However, the value of $\tilde{\omega}^-$ is unaffected by the
presence of a large slip length when $\eta_1 \gg \eta_2$.  
Thus, we can see that, like the single layer case,
a double layer of Newtonian fluids does not exhibit a $q$-independent decay rate,
even in the presence of liquid-liquid slip.

Examining the normal mode amplitude ratio as a function of wave
number, shown in the inset of Fig.~\ref{fig:NewtDL}, we see that at the
upper surface the fast mode amplitude always dominates, while  at the buried
interface the slow mode always dominates. Thus, the
decay of the height-height correlation functions, which are
given by Eq.~(\ref{eq:HeightCorrDL}), are essentially of the form of a
single exponential decay, $S_1 (q, t)\sim e^{-\tilde{\omega}^- t}$
and $S_2 (q, t) \sim e^{-\tilde{\omega}^+ t}$.  The large size of the amplitude ratios 
at the two surfaces is  due primarily to the large viscosity difference between 
the two layers. If we were to consider a two layer system having similar viscosities and 
similar interfacial tensions at the two surfaces, we would find amplitude ratios at the
two surfaces of order unity.  Of course, the rates themselves would be approximately
equal as well in this limit.  

\subsection{One Maxwell Fluid, One Newtonian Fluid}
\label{two-fluid:one-max}

\begin{figure*}
\includegraphics[width=8.5cm]{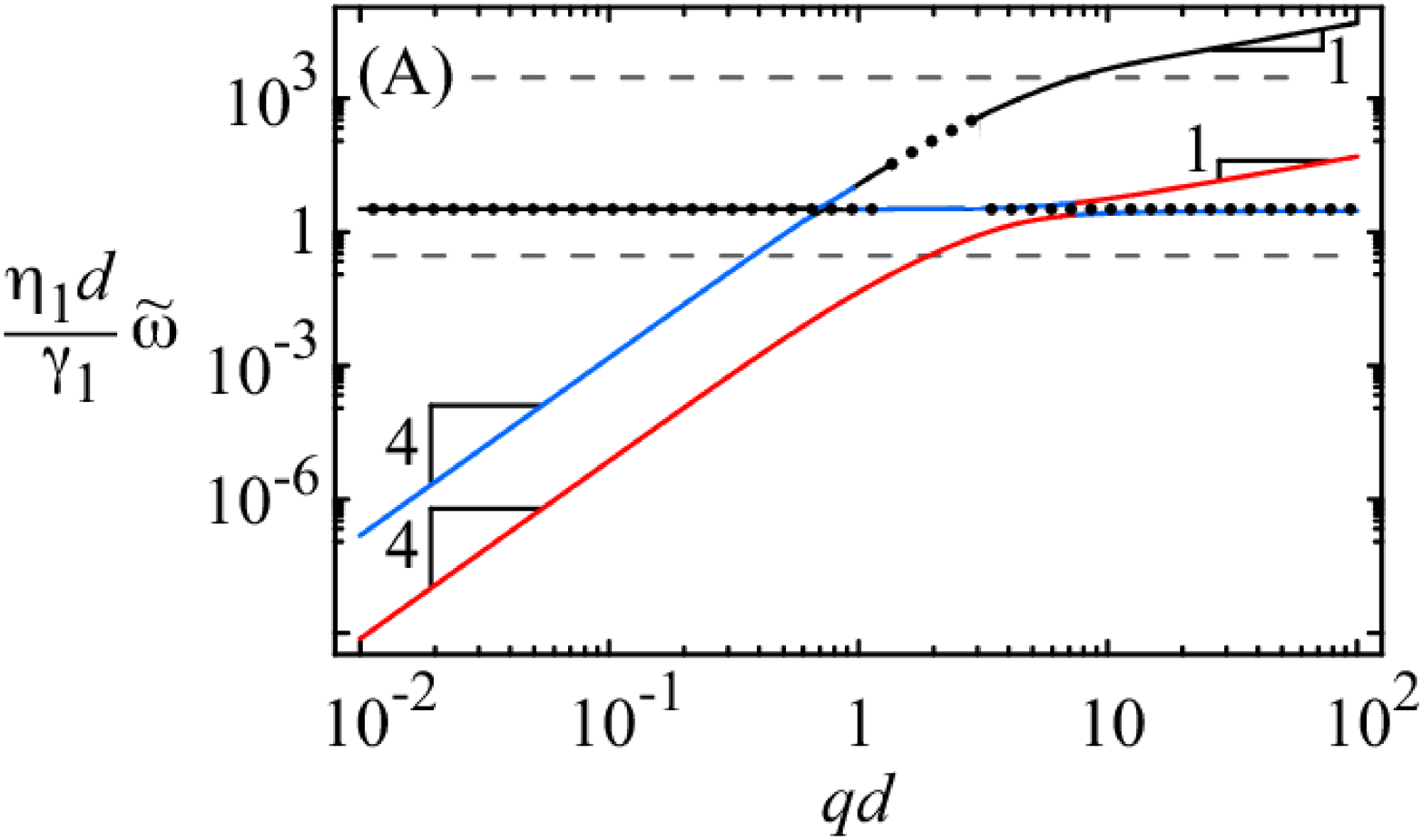}
\hspace{13pt}
\includegraphics[width=8.5cm]{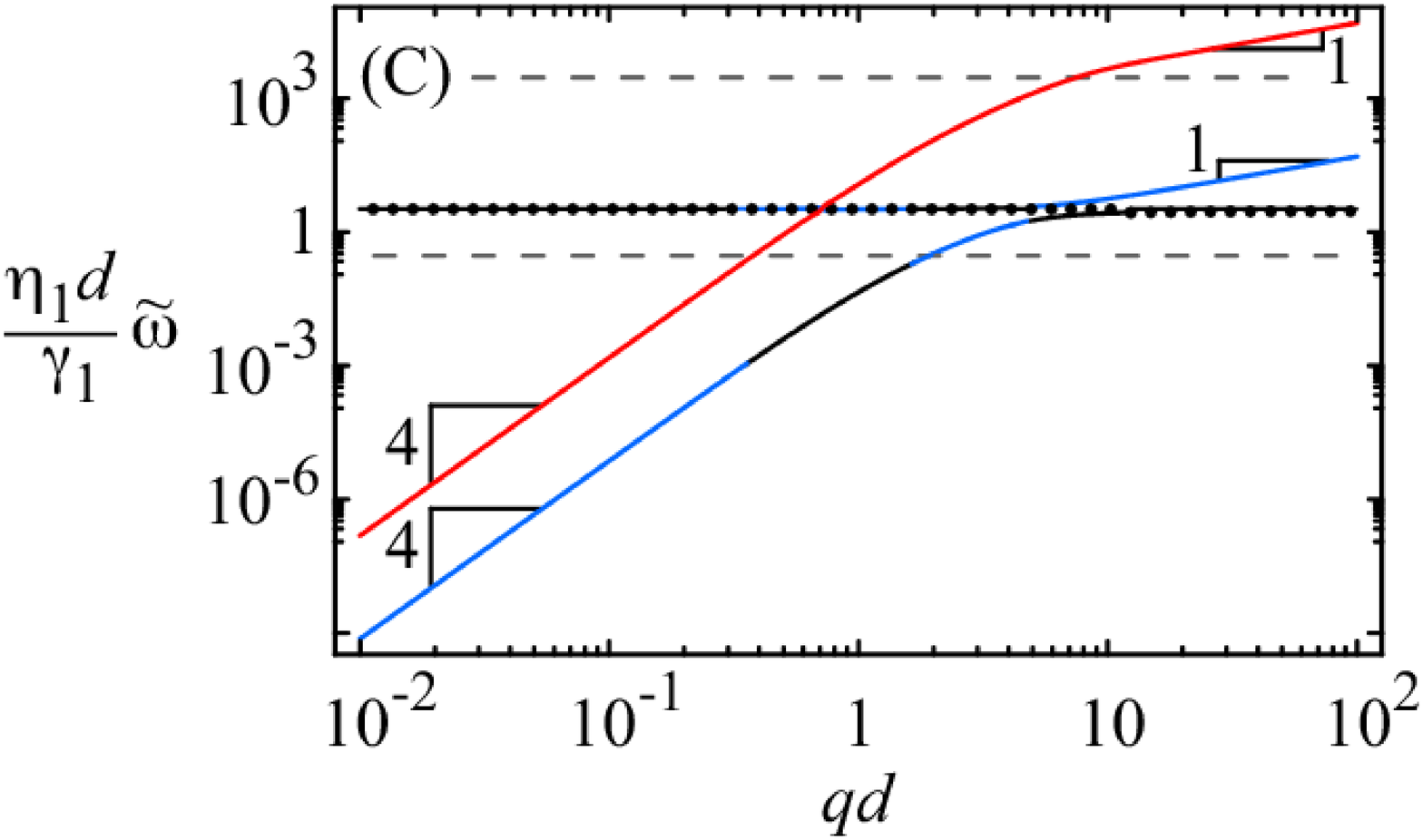}
\includegraphics[width=8.5cm]{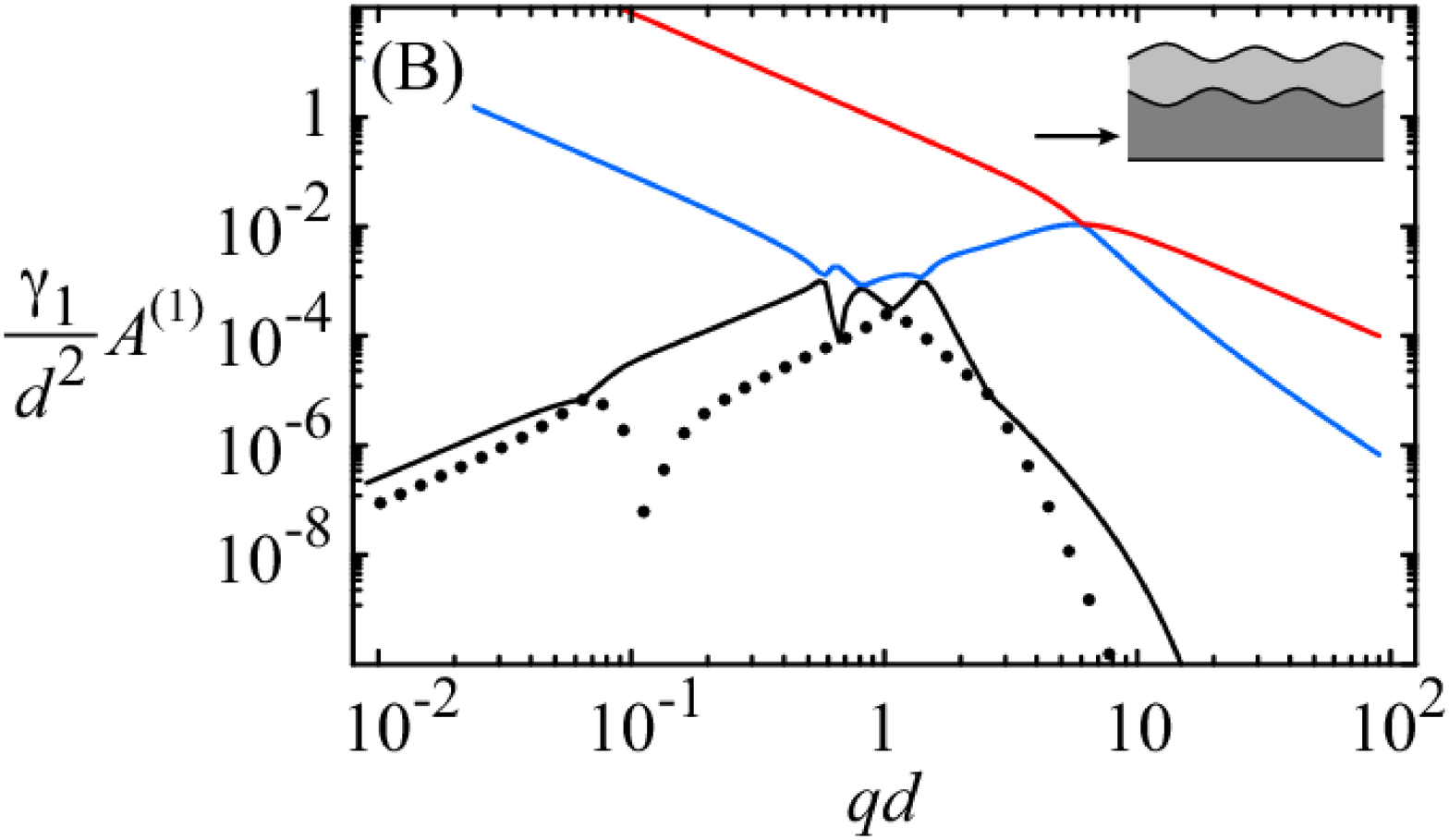}
\hspace{13pt}
\includegraphics[width=8.5cm]{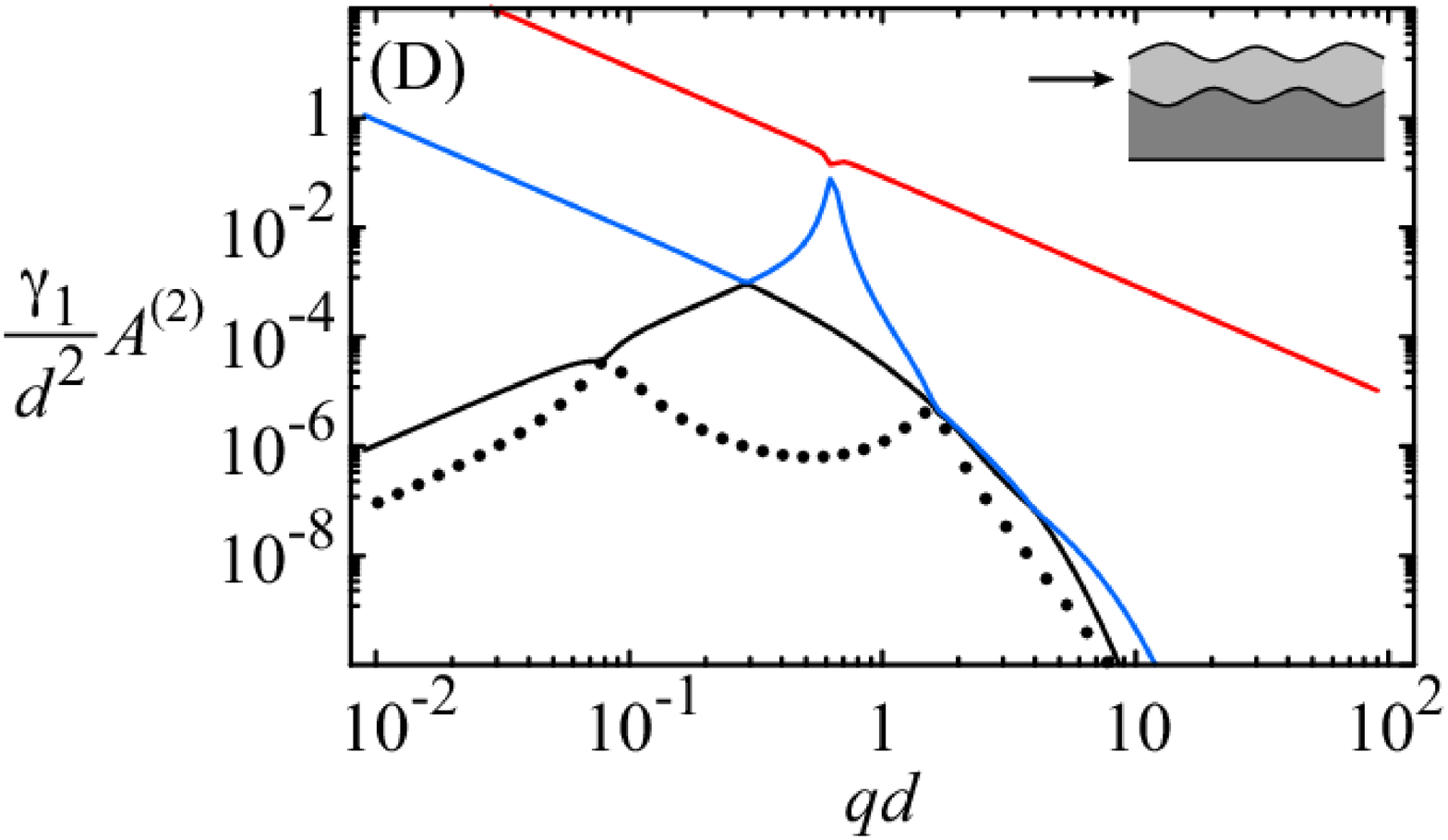}
\caption{\label{fig:MaxwellDL} (\textit{color online}) Dimensionless
decay rates $\eta_1d \tilde{\omega}^\pm /\gamma_1$ (\textit{A, C})
and amplitudes $\gamma_1 A/d^2$ at the buried (\textit{B}) and
upper (\textit{D}) interfaces, as a function of $q d$ for a double layer
 system with a buried Maxwell PBrS layer and an upper Newtonian
 PS layer in  the absence of slip,  using the parameter values given
 in Table~\ref{tab:values}. (\textit{A, C}) The rates shown in these two
 figures are identical, except that the line styles in Fig. A correspond
 to those in Fig. B, while the line styles in Fig. C correspond
to those in Fig. D.  The dashed lines indicate
the approximate window of decay rates that can be measured in the
experiments, $10^{-3} s^{-1} < \tilde{\omega}<10 s^{-1}$.  (\textit{B, D})
The amplitudes in these figures are indicated by dotted, black, blue, and red lines,
in order of increasing amplitude.}
\end{figure*}

When one of the fluids in the double layer is viscoelastic, we expect
to find additional decay rates, as we found for the single viscoelastic layer.
Here, we focus on the case of a buried viscoelastic fluid.  It is known from
the single layer experiments that PS fluid layers -- the
upper fluid in the double layer experiments -- are well
described by a purely viscous Newtonian fluid model~\cite{kim:03}.
On the other hand, we expect that the PBrS layer to have longer stress
relaxation times. We model this viscoelastic layer using the Maxwell model,
so that $\eta_1 (\omega)$ is of the form of Eq.~(\ref{eq:EtaMaxwell}). This
approach introduces two unknown rheological parameters for the material
(namely, $E$ and $\tau$), which we discuss more fully below. For simplicity,
we analyze this system with stick boundary conditions at all interfaces,
\emph{i.e.} $\lambda=0$.

From examining the zeros of  Eq.~(\ref{eq:DetSigma})
we find four separate decay rates for this dynamical system, in agreement
with the arguments given in Appendix~\ref{app:NumFreq}.
These decay rates for a double layer system with a buried PBrS layer
and an upper PS layer are shown in Figs.~\ref{fig:MaxwellDL}A and
~\ref{fig:MaxwellDL}C, using the parameters given in Table~\ref{tab:values} 
(these two plots are identical, except that the line styles in each are chosen
to correspond to Figs.~\ref{fig:MaxwellDL}B and
~\ref{fig:MaxwellDL}D, respectively; we explain this in more detail below).
The behavior of the decay rates is qualitatively similar to the single layer
viscoelastic case. In particular, each rate has two major
scaling regions: a $q$-dependent region, where the decay rate scales
like the corresponding one in the Newtonian fluids case (i.e. $\sim q^4$ for
small $q d$ or $\sim q$ for large $q d$), and a $q$-independent
region. In the single layer case, however, the crossover between
these two regions is abrupt and occurs at $q d \sim {\mathcal O}(1)$. This is
not the case for the double layer system; indeed, one decay
rate is essentially constant over the entire range of wave numbers shown in
Fig.~\ref{fig:MaxwellDL}, with its $q$-dependent regime not appearing until
higher wave numbers (data not shown).

In principle, all four decay rates shown in Fig.~\ref{fig:MaxwellDL} contribute to
the height-height correlation functions at both the upper and buried interface.
As in the single layer case, though, the amplitudes corresponding to each
decay rate are important in determining which rates can be observed
experimentally in these correlation functions.   Figs.~\ref{fig:MaxwellDL}B and
~\ref{fig:MaxwellDL}D show the amplitudes for each decay rate at the buried and
upper interfaces, respectively.  The decay rates shown in Figs.~\ref{fig:MaxwellDL}A
and~\ref{fig:MaxwellDL}C are plotted using the same line
style (dotted, black, blue, and red) as their corresponding amplitude at the buried
and upper interfaces, respectively.  In these two figures we represent the experimentally
accessible range of timescales by the region between the two grey dashed lines.
From examining Figs.~\ref{fig:MaxwellDL}B and~\ref{fig:MaxwellDL}D it is clear that
the ratios of the various amplitudes vary by many orders of magnitude over the experimentally
accessible wave number range.  Also, the dominant amplitude  at both
interfaces is always $q$-dependent, as was seen in the single layer case.

We consider first the dynamics of buried interface shown in 
Figs.~\ref{fig:MaxwellDL}A and~\ref{fig:MaxwellDL}B.
If we focus on the wave number regime accessible in the experiments
of Lal \textit{et al.}, $0.8 \lesssim q d \lesssim 2$, we see that the slow, $q$-dependent
rate, which has the largest amplitude at the buried interface, is too slow to be
detected experimentally.  The other decay rates all have amplitudes of approximately the same
magnitude in this region. Thus, it appears that the buried layer dynamics
should, in fact, be best described by a height autocorrelation
function having multiple exponential decays.   The data of Lal \textit{et. al}, however,
was consistent with single exponential decays at this interface.  This could
be due to the fact that decay rates do not differ substantially in this region (see the
blue, black, and dotted curves in Fig.~\ref{fig:MaxwellDL}A), thus rendering it
difficult to observe the multi-exponential decay profiles.  We do predict, however, 
that the $q$-independent decay rate should be observable at the buried layer,
in agreement with the experimental data.

We now turn to the dynamics of the free surface shown in 
Figs.~\ref{fig:MaxwellDL}C and \ref{fig:MaxwellDL}D.  We see that for
wave numbers $ q d \simeq 0.7$, it should be possible to observe a double exponential
decay, since the amplitudes of the dominant, $q$-dependent decay rate (red line)
and subdominant $q$-independent rate (blue line) are approximately equal, as
shown in Fig.~\ref{fig:MaxwellDL}D. Furthermore, we can see from
Fig.~\ref{fig:MaxwellDL}C that the $q$-dependent rate is slower than the
$q$-independent rate in the experimental range of wave numbers,
$0.8 \lesssim q d \lesssim 2$.  The experiments do indeed observe a double
exponential decay with a faster $q$-dependent rate and a slower $q$-independent
rate; however, the amplitude of the $q$-independent decay is larger than that
of the $q$-dependent decay in the experiments, whereas the opposite is true
for our theoretical predictions.

We suspect that pursuing more detailed numerical fits of the theory to 
the current experiments is unproductive
due primarily to the inadequacies of the Maxwell model. In particular, a more
realistic description of the stress
relaxation in the PBrS layer will allow for a spectrum of relaxation times, leading to a band
of decay rates at a given wave number. As is clear from Figs.~\ref{fig:MaxwellDL}C and D,
the amplitudes of the decay rates are highly nonlinear, and thus are sensitive to the details of the
viscoelastic model used for the fluid.  As the theory
currently stands, it appears to be fortuitous that one may observe the
double exponential decay at the free surface, since
amplitude ratio between the two dominant modes (red and blue lines)
in this figure have an extremely narrow
maximum at  $q d \approx 0.7$.  We expect, however, that a broadening of
the relaxation time spectrum in the material will increase the width of the features in
Fig.~\ref{fig:MaxwellDL}D, making the observation of
the multiple exponential decay in the data much more plausible.
It is important to note that the use of a more realistic viscoelastic model for the
buried fluid should not affect the existence of the $q$-independent decay rates themselves.
Indeed, the scaling arguments
given in section~\ref{sec:scaling} show that the presence of the $q$-independent
decay rate is a robust feature of the elastic response of the fluid. Thus, a more realistic model
for the buried fluid should be able to account for the discrepancies between our
theory and the experimental results of Lal \textit{et al.}

Finally, we note that when the upper fluid is a Maxwell fluid and the lower fluid is a
Newtonian fluid, an additional decay rate appears, in agreement with the arguments
given in Appendix~\ref{app:NumFreq}.   Although four of the
rates behave in a manner similar to that of the four decay
rates for a buried Maxwell fluid -- with each rate
displaying one region of $q$-independent behavior and another region
of $q$-dependent, Newtonian-like behavior -- the fifth rate is
essentially constant over the wave numbers of interest.  Furthermore, we find that
at certain wave numbers the amplitude of one of the $q$-independent decay rates
can actually be as large as the largest $q$-dependent amplitudes in the height-height
correlation functions, at both the buried and upper interfaces (data not shown).
Thus, our analysis suggests that detecting the
$q$-independent rates may be even easier when the upper fluid
is viscoelastic.  It is important to repeat, however, that we do not suspect that it is
a viscoelastic response of the upper PS layer that
is leading to the $q$-independent decay rates seen in the experiments,
since the purely viscous model fits the single layer PS data well~\cite{kim:03}.

\section{Conclusion}
\label{sec:conclusion}

In this article we have examined in detail the overdamped dynamics
of purely viscous and viscoelastic supported films, taking the effects of both
liquid/substrate and liquid/liquid slip into account. These calculations show that a simple
approach to understanding the appearance of wave number independent decay rates in the height
autocorrelation function $S(q,t)$ in both single and double layer films is found by allowing for a
viscoelastic response of the material. Simple scaling arguments show that a viscoelastic material will
exhibit such dynamics over some finite wave number range. The more detailed continuum mechanics
calculations based on the Maxwell fluid model presented above show that this
wave number independent scaling regime is experimentally accessible, at least for
a viscoelastic material with the appropriate values of  the plateau modulus $E$ and the stress
relaxation time $\tau$.

More generally, the appearance
of a $q$-independent relaxation rate in $S(q,t)$ appears to be a signal of a viscoelastic
response of the supported polymer film on the time scales probed by the measurement. The observed
decay time is strongly controlled by the longest stress relaxation time in the material, and thus
may serve as an important measure of the rheological properties of such thin supported films.
In the numerical results that we presented
our choice of the stress relaxation time $\tau$ is
constrained by the value of the $q$-independent decay time
measured in the experiments, which is of the order of $100$s.
In order to fit the data,
we had to take $\tau$ to be of this same order, which is almost two orders
of magnitude larger longer
than might be expected for PS at the reported molecular weight~\cite{pedersen:99}. We
suggest that discrepancy may
be attributed to two factors. First, the bromination of the PS
changes the microscopic dynamics
of dynamics of the polymer. Second, the narrow thickness of the
PBrS layer (the layer is
about three radii of gyration of the chain in $\theta$-conditions)
adds further constraints to the chain dynamics leading to a longer
stress relaxation time.

Given a determination of $\tau$, we found that our choice of the plateau
modulus was not as precisely constrained by the data.  If we choose
$E \tau/\eta_1 \gg 1$, the value of the $q$-independent decay time
is decreased below $\tau$.  Since the value of $\tau$ needed to fit the data
is already anomalously large, we wish to choose the smallest possible value of $\tau$:
this restricts us to the region $E \tau/\eta_1 \ll 1$.
As long as we choose a value of $E$ within this region, though,
the value of the $q$-independent decay rate is unaffected by it.
Rather, the principal effect of varying the plateau modulus is to shift the
various amplitude ratios of the multiple decay rates, in particular the
wave numbers at which these ratios approach unity.

This points out two more general
results. The first is that the appearance of multiple exponential 
decay rates of the height autocorrelation
function $S(q,t)$ for a single layer system is a generic feature 
of our viscoelastic model. In a two-layer system, even purely viscous
fluids admit a double exponential decay. If either of the layers is viscoelastic,
then this double decay becomes a more complicated multiple decay as discussed above.
Based on these calculations we
suspect that the observation of multiple exponential
decays of $S(q,t)$ in a single layer system may be taken as evidence of
the viscoelasticity of the supported film. Secondly,
our calculations show that observable multiple exponential
decays depend on amplitude ratios that are in turn
controlled by the plateau modulus of the material. These ratios
can vary by many orders of magnitude with the plateau modulus and wave number. The
observation of multiple exponential decays requires the ratio of the two largest
amplitudes to be near unity;
from our Maxwell-model based calculations such occurrences occupy a
small part of the phase space spanned by
wave number and plateau modulus. While we expect that a broader
spectrum of stress relaxation times in the material
will enlarge the region of phase space over which these multiple
exponential decays can be seen, such multiple
exponential decays may not be a generically observable feature
of viscoelastic supported films.  When such multiple exponential decays are observed,
however, they should provide a sensitive window onto the plateau modulus
of the material and thus measure the entanglement length in the layer.

Finally, we point out that these calculations do not in general exclude other potential mechanisms
as the underlying cause
of the dynamics as observed by XPCS. They do show, however, that these data are not consistent
with the dynamics of two immiscible Newtonian fluids either with
stick or slip boundary conditions
at their boundaries. Moreover, one can show that postulating more complex surface energy
functionals -- including, for example, an interfacial bending modulus --
will only lead to relaxation rates of $S(q,t)$ that are even more strongly
dependent on the wave number, which would be inconsistent
with the data.  Therefore, such considerations can also be excluded.  It appears that
most simple way to account for the data is to postulate a viscoelastic response of
the supported film with a stress relaxation time on the order of the observed decay rate of
the height autocorrelation function.

A more detailed analysis of the
interfacial dynamics of complex fluids via XPCS holds the
promise of probing molecular motion in confined geometries that may be interpreted
with the aid of the continuum modeling presented in this article. If our calculations
are to serve in this manner, however, they must first be tested further
by experiment on rheologically well-characterized materials of thickness large
enough to discount the effects of molecular confinement.

\section*{Acknowledgements}

MLD and AJL thank J. Lal for providing unpublished data and
for enjoyable discussions. MDL and AJL were supported in part by
NASA NRA 02-OBPR-03-C.

\appendix
\section{Solution of the Time-dependent Stokes Equation}
\label{app:NavierStokes}

For a low Reynolds number fluid whose velocity and pressure are of
the form of Eq.~(\ref{eq:Transform}), the time-dependent Stokes equation is
given by
\begin{align}
\label{eq:NavierStokes}
i \omega \rho \, {\bf v} (q, z, \omega)
e^{i q x} =&- \eta (\omega) {\bf \nabla}^2 \left[{\bf v} (q, z, \omega) e^{i q x}\right] \\
\notag
&+ {\bf \nabla} \left[P (q, z, \omega) e^{i q x}\right],
\end{align}
where $\rho$ is the fluid density and $\eta (\omega)$ is the
frequency-dependent fluid viscosity.  We also assume that the fluid
is incompressible,
\begin{equation}
{\bf \nabla} \cdot {\bf v}(x,t)=0 \quad \Rightarrow \quad v_x
(q, z, \omega) = \frac{i}{q} v_z' (q, z, \omega),
\end{equation}
where the prime indicates a derivative with respect to the
$z$-coordinate.  Using the identity ${\bf \nabla} \times
\left[{\bf \nabla} \psi(x,z,t) \right] = 0$ (for any scalar
function $\psi$), we can take the curl of  Eq.~(\ref{eq:NavierStokes})
 to eliminate the pressure,
\begin{equation}
\label{eq:OmegaDE} i \omega \rho \, \Omega_y (q, z, \omega) + \eta
(\omega) \left[ \partial_z^2-q^2\right] \Omega_y (q, z, \omega) = 0,
\end{equation}
where ${\bf \Omega} (x,z,t) \equiv {\bf \nabla} \times {\bf v}
(x, z,t)=\Omega_y (x,z,t) \hat{y}$ is the fluid vorticity,
\begin{equation}
\label{eq:OmegaDef} \Omega_y (q, z, \omega) = \frac{i}{q} v_z'' (q,
z, \omega) - i q v_z (q, z, \omega).
\end{equation}
The solution to Eq.~(\ref{eq:OmegaDE}) can be written as
\begin{equation}
\Omega_y (q, z, \omega) = 2 q C_1 \cosh (k z) + 2 q C_2 \sinh(k z),
\end{equation}
where $C_1$ and $C_2$ are constants of integration and $k \equiv
\sqrt{q^2 - \frac{i \omega \rho}{\eta (\omega)}}$.  The
$z$-component of the velocity obeys the differential equation
Eq.~(\ref{eq:OmegaDef}), which has the solution
\begin{align}
v_z (q, z, \omega) =&  C_3 \cosh (q z)+ C_4 \sinh (q z) \\
\notag &+ \int_0^\infty dz' G(z-z') \Omega_y (q, z', \omega),
\end{align}
where the first two terms are the homogeneous solution to
Eq.~(\ref{eq:OmegaDef}), and $G(z)$ is the Green's function for the
operator $i ( \partial_z^2-q^2)/q$, $G(z) = -i \sinh(q z) \theta
(z)$, $\theta (z)$ being the Heaviside step function.  Then the
velocity is given by
\begin{align}
\label{eq:VelocityI}
&v_z (q, z, \omega) = 2 C_1' q^2 \cosh (k z) +
2 C_2' q^2 \sinh(k z) \\
\notag
&+\left(C_3 -2 C_1' q^2 \right) \cosh (q z) +
\left(C_4 -2 C_2' k q \right) \sinh (q z),
\end{align}
where $C_j' \equiv C_j/(k^2-q^2)$ for $j=1,2$.

The pressure can be obtained from the $x$-component of  Eq.~(\ref{eq:NavierStokes}),
\begin{equation}
\label{eq:Pressure} P (q, z, \omega) = \frac{\eta (\omega)}{q^2}
\left[ v_z''' (q, z, \omega)-k^2 v_z' (q, z, \omega)\right].
\end{equation}
Note that the $z$-component of Eq.~(\ref{eq:NavierStokes})
is automatically satisfied by this pressure and the velocity field
given by Eq.~(\ref{eq:VelocityI}).

For overdamped fluids, where the inertial term in the time dependent Stokes
equation -- that is, the left-hand side of
Eq.~(\ref{eq:NavierStokes}) -- is negligible, we can set $\rho = 0$,
which causes $k \rightarrow q$.  In this case, the velocity
Eq.~(\ref{eq:VelocityI}) is given by
\begin{align}
\label{eq:VelocityNI}
v_z (q, z, \omega) = &C_3\cosh (q z) + (C_4-C_2) \sinh(q z) \\
\notag
&+ C_1 q z \sinh (q z) +C_2 q  z \cosh (q z).
\end{align}

\section{The Fluctuation-Dissipation Theorem}
\label{app:FDT}

The fluctuation-dissipation theorem relates correlations between
 the orthogonal dynamical variables describing some system at
 equilibrium to the response of these variables to external forces.
 For the single fluid layer system we consider in this paper, there
 is only one dynamical variable, the height of the fluid,
 $h (x, t)$. For the double layer
 system, there are two orthogonal dynamical
variables, the normal modes amplitudes $h_\alpha (x, t)$,
where $\alpha = \pm$.  In both cases,
the response function can be obtained from the normal stress boundary
condition(s) at the fluid interface(s), which in the normal mode
basis can be written in the form
\bb
\label{eq:XiDef}
\frac{h_\alpha (q, \omega)}{\xi_\alpha (q, \omega)} =
f_\alpha (q, \omega).
\en
The fluctuation-dissipation theorem relates the response function
$\xi_\alpha (q, \omega)$ to the height-height correlation function
\begin{align}
\label{eq:HeightCorr1}
S_\alpha (q, t) \equiv & \frac{\left(2\pi\right)^3}{k_B T L^2}
\left<h_\alpha (q, t) h_\alpha^* (q, 0) \right> \\
\notag
=&\int_{-\infty}^\infty
\frac{d \omega}{2 \pi i \omega} e^{-i \omega t}
\left[\xi_\alpha (q, \omega) - \xi_\alpha^*(-q, \omega)\right].
\end{align}
where $L^2 \equiv \int d^2 x$ and the star indicates complex conjugation.

The integral in Eq.~(\ref{eq:HeightCorr1}) can be evaluated by
contour integration using a semi-circular, negatively
oriented contour $\mathcal{C}$ whose arc lies in the negative
imaginary half-plane and whose diameter is the real axis. We can see
from Eq.~(\ref{eq:XiDef}) that the frequencies for which
$\xi^{-1}_\alpha (q, \omega) = 0$ are precisely the normal mode
frequencies $\omega_n$ [i.e. the frequencies that satisfy the normal
stress boundary conditions Eq.~(\ref{eq:XiDef}) for $f_\alpha =0$].  
Because the system is overdamped, these
frequencies lie on the negative imaginary axis, within the contour
$\mathcal{C}$; therefore, each normal mode frequency contributes a
non-zero residue to the contour integral.   Furthermore, it can be
shown that, for all systems we consider in this paper,
$\xi_\alpha^*(-q, \omega) = \xi_\alpha (q, -\omega)$.  This implies that
the second term in Eq.~(\ref{eq:HeightCorr1}) has no poles inside the
contour, nor does it contribute to the residues at the poles
$\omega_n$.  Assuming the poles at each $\omega_n$ are simple poles
(which can be shown to be true in all cases),
Eq.~(\ref{eq:HeightCorr1}) becomes
\begin{equation}
\label{eq:HeightCorr}
S_\alpha (q, t) = -\sum_{\omega_n}
\lim_{\omega \rightarrow \omega_n} \left[\left(\omega-
\omega_n\right) \frac{\xi_\alpha (q, \omega)}{\omega} e^{- i \omega
t}\right].
\end{equation}

\section{Number of Normal Mode Decay Rates}
\label{app:NumFreq}

The number of decay rates in a dynamical system is simply related to
the number of independent degrees of freedom~\cite{lubensky:book}.
It is therefore somewhat counterintuitive to observe two decay rates
for capillary waves on a single supported viscoelastic layer and either
four or five decay rates
in the double layer system. It is perhaps most
surprising to find the number of decay rates to change upon
inverting the order of the layering of the two materials. In this
appendix we explain this result in a simple way by considering the
equations of motion in the time domain rather than in the frequency
domain, as we have done throughout the rest of this article.

Let us first consider the single layer case.
 Here and throughout this appendix we suppress any
wave number $q$ dependence for clarity.  From the boundary
condition Eq.~(\ref{eq:BC3SL}) we see that $v_z (z, \omega) \propto
- i \omega h(\omega)$. Using the boundary conditions
Eqs.~(\ref{eq:BC1SL})-(\ref{eq:BC4SL}), we write the velocity in the
form
\begin{equation}
v_z (z, \omega) = - i \omega
h(\omega) \psi (z).
\end{equation}
Taken in combination with the stress boundary condition
Eq.~(\ref{eq:StressBalanceSL1}) we find
\begin{equation}
\label{eq:hOmegaSL}
\gamma q^2 h(\omega) = - i \omega \eta(\omega) h(\omega)
f(0),
\end{equation}
where $f(z) \equiv 3 \psi'(z)-\psi'''(z)/q^2$.

For a Newtonian fluid, $\eta (\omega) = \eta$.  Converting
Eq.~(\ref{eq:hOmegaSL}) back to the time domain we
recover a first order differential equation for the decay
 of the surface height field:
\begin{equation}
\gamma q^2 h(t) = \eta f(0) \dot{h}(t).
\end{equation}
 Thus, we expect only one decay rate at a given wave number
 for capillary waves on a single supported Newtonian fluid.

For the single supported Maxwell fluid the situation is more
complicated. Now the viscosity is a complex frequency dependent
function, which is given by Eq.~(\ref{eq:EtaMaxwell}). As a result,
conversion of Eq.~(\ref{eq:hOmegaSL}) into the time domain
 yields an integro-differential equation:
\begin{equation}
\label{eq:hODEMaxwell1}
\frac{\gamma q^2}{f(0)} h(t) = \eta_0
\dot{h}(t) - \frac{E_0}{\tau} \int_{-\infty}^t \!\!\!\! dt' h(t')
e^{-\frac{t-t'}{\tau}}.
\end{equation}
The time evolution of the surface height now depends on the entire
deformation history of the surface convoluted with an exponential
memory kernel. The simple form of the memory kernel is an artifact
of the simplicity of the Maxwell fluid, but the general structure of
this equation of motion holds for any more complex model of
viscoelasticity.

Taking a time derivative of Eq.~(\ref{eq:hODEMaxwell1}) we may write
a single, second order differential equation for the amplitude of
the height field at wave vector ${\bf q}$:
\begin{equation}
\frac{\gamma q^2}{f(0)}\left[\tau \dot{h}(t)+h(t)\right]=
\eta_0\left[ \tau \ddot{h} (t)+\dot{h}(t)\right] -E_0 h(t).
\end{equation}
From this second order differential equation, which has two
independent exponentially decaying solutions, we know to expect a
double exponential decay for capillary waves on the supported
Maxwell fluid.

We now consider the case of two fluid layers. As in the single layer
case we can use the boundary conditions
Eqs.~(\ref{eq:BC1DL})-(\ref{eq:BC4DL}) to write the fluid velocities
in the form
\begin{equation}
v_{k,z} (z, \omega) = - i \omega \left[
h_1 (\omega) \psi_{k1} (z)+ h_2 (\omega) \psi_{k2} (z)\right],
\end{equation}
with $k=1,2$ indexing the fluid layers. There remain three more
boundary conditions. One enforces tangential stress continuity at
the fluid/fluid interface; the other two set the difference in the
normal stresses at both the fluid/fluid interface and the free
surface in terms of their respective surface tensions. These
conditions may be written as, respectively, 
\begin{equation}
\label{eq:hOmegaDL1} \eta_1 (\omega) \sum_{k=1,2} g_{1k}(0) h_k
(\omega) = \eta_2 (\omega)\sum_{k=1,2} g_{2k}(0) h_k (\omega),
\end{equation}
and
\begin{align}
\label{eq:hOmegaDL2}
\gamma_1 q^2 h_1 (\omega) &= \sum_{j,k=1,2} (-1)^j i
\omega h_k (\omega) \eta_j (\omega) f_{jk}(d_2) \\
\label{eq:hOmegaDL3}
\gamma_2 q^2 h_2 (\omega) &= \sum_{k=1,2} i
\omega \eta_2 (\omega) g_{2k}(0) h_k (\omega),
\end{align}
where $f_{jk}(z) = 3 \psi_{jk}' (z)-\psi_{jk}'''(z)/q^2$ and
$g_{jk}(z) = q^2 \psi_{jk} (z)+ \psi_{jk}''(z)$.  In a manner
analogous to the one used above with the single layer problem, we can
rewrite Eqs.~(\ref{eq:hOmegaDL1})-(\ref{eq:hOmegaDL3}) in the time
domain. Note that the presence of the $i \omega$ factor in
Eqs.~(\ref{eq:hOmegaDL2}) and~(\ref{eq:hOmegaDL3}) implies a time derivative.

When both fluids are Newtonian Eq.~(\ref{eq:hOmegaDL1}) involves no
time derivatives, whereas both Eqs~(\ref{eq:hOmegaDL2})
and~(\ref{eq:hOmegaDL3}) are first order differential equations.
Having a system of two first order differential equations, we expect
two normal mode decay rates for a double layer of two
immiscible Newtonian fluids.

If either the upper or buried fluid is a Maxwell fluid, then it
clear from the arguments given in the single layer case that
Eq.~(\ref{eq:hOmegaDL1}) becomes a first-order differential equation
in the time domain, whereas Eq.~(\ref{eq:hOmegaDL2}) becomes a
second order ODE. Eq.~(\ref{eq:hOmegaDL3}), on the other hand,
depends only on the viscosity of the upper fluid, $\eta_2$, so it
becomes a second order differential equation only when the upper
fluid is a Maxwell fluid; otherwise, it is a first order differential equation.
Therefore, if the upper layer is a
Newtonian fluid we have a set of two first order differential
equations and one second order differential equation; we expect
there to be \textit{four} decay rates in this case. If the
upper fluid is a Maxwell fluid and the lower one is Newtonian,
however, we now have a dynamical system described by one first
order differential equation and two second order ones.
In this case there will be
\textit{five} decay rates.

 \end{document}